\documentclass[a4paper,aps,prl,superscriptaddress,reprint,longbibliography]{revtex4-1}

\usepackage[utf8]{inputenc}
\usepackage[english]{babel}
\usepackage{amsmath,amssymb}
\usepackage{graphicx}
\usepackage{color}
\usepackage{bm}

\DeclareMathOperator{\diag}{\mathrm{diag}}

\begin{document}
	
\title{Spin-wave chirality and its manifestations in antiferromagnets}

\author{Igor \surname{Proskurin}}
\email{iprosk@ouj.ac.jp}
\affiliation{Division of Natural and Environmental Sciences, 
	The Open University of Japan,
	Chiba 261-8586, Japan}
\affiliation{Institute of Natural Sciences, Ural Federal University, Ekaterinburg 620002, Russia}

\author{Robert~L. \surname{Stamps}}
\affiliation{School of Physics and Astronomy, University of Glasgow, Glasgow, UK, G12 8QQ}

\author{Alexander~S. \surname{Ovchinnikov}}
\affiliation{Institute of Natural Sciences, Ural Federal University, Ekaterinburg 620002, Russia}
\affiliation{Institute for Metal Physics, RAS, 620137, Ekaterinburg, Russia}

\author{Jun-ichiro \surname{Kishine}}
\affiliation{Division of Natural and Environmental Sciences, The Open University of Japan, 	Chiba 261-8586, Japan}

\begin{abstract}
As first demonstrated by Tang and Cohen in chiral optics, the asymmetry in the rate of electromagnetic energy absorption between left and right enantiomers is determined by an optical chirality density \cite{Tang2010}.  Here, we demonstrate that this effect can exist in magnetic spin systems.  By constructing a formal analogy with electrodynamics, we show that in antiferromagnets with broken chiral symmetry the asymmetry in local spin-wave energy absorption is proportional to a spin-wave chirality density, which is a direct counterpart of optical zilch.  We propose that injection of a pure spin current into an antiferromagnet may serve as a chiral symmetry breaking mechanism, since its effect in the spin-wave approximation can be expressed in terms of additional Lifshitz invariants. We use linear response theory to show that the spin current induces a nonequilibrium spin-wave chirality density.
\end{abstract}

\maketitle


\textit{Introduction.} --- Chirality describes mirror image symmetry or the lack thereof. \cite{Kelvin1904}.  Circularly polarized light provides a simple example. It has been known for a long time that when circularly polarized light interacts with a chiral molecule, the excitation rate is different for left and right polarizations, leading to remarkable effects such as natural optical activity and circular dichroism \cite{Barron2004}.  After Lipkin's discovery of a chirality conservation law for the electromagnetic Maxwell's equations \cite{Lipkin1964}, it was realized that the electromagnetic field can be characterized by a locally conserving chirality density that is odd under spatial inversion ($\mathcal{P}$) and even under time reversal ($\mathcal{T}$) transformations. These symmetry properties are consistent with Barron's proposal of \emph{true chirality} \cite{Barron1986} that generalized the original definition by Kelvin \cite{Kelvin1904}.

Tang and Cohen realized that in local light-matter interactions of structured electromagnetic fields inside materials with broken chiral symmetry, electromagnetic chirality determines the asymmetry in the electromagnetic energy absorption rate \cite{Tang2010}.
Later, Bliokh and Nori demonstrated that chirality density in \cite{Tang2010} is directly related to polarization helicity and energy density \cite{Bliokh2011}.  Further progress in understanding mutual relations between optical helicity, duality symmetry, and spin angular momentum of light was developed in Refs.~\cite{Coles2012,Cameron2012,Bliokh2013,Bliokh2014}.
These discoveries paved the way for possible practical applications of chiral electromagnetic fields in optics and plasmonics \cite{Hendry2010,Tang2011,Hendry2012,Canaguier2013}. The purpose of  this Letter is to demonstrate that this effect can be found in some magnetic spin systems. We consider the example of an antiferromagnetic material whose magnetic excitations -- know as spin waves -- can display some key properties analogous to optical light \cite{Akhiezer1968}.

Spin dynamics in antiferromagnets attracted considerable attention recently \cite{Nunez2006,Haney2008,Gomonay2010,Hals2011,Swaving2011,Cheng2012,Gomonay2012,Tveten2013,Takei2014,Cheng2014,Cheng2014a,Takei2015,Yamane2016,Khymyn2016,Rezende2016,Rezende2016a,Daniels2015} from the perspective of spintronics \cite{Jungwirth2016}. In this respect, non-centrosymmetric antiferromagnets are especially interesting. Lack of the inversion symmetry lifts the degeneracy between left and right polarized spin waves inside such materials making possible the observation of magnonic Nernst effects \cite{Cheng2016a,Zyuzin2016} or development of spin-wave field effect transistor devices \cite{Cheng2016}. Recently, an antiferromagnetic version of a chiral magnetic effect was proposed \cite{Sekine2016} thus establishing a link between antiferromagnets and Weyl semimetals \cite{Zyuzin2012}.

In this Letter, we examine the dynamics of antiferromagnetic spin-wave excitations and draw analogy with electrodynamics. This allows us to generalize the method of nongeometric symmetries, originally developed for the free electromagnetic field \cite{Fushchich1987}, to antiferromagnetic spin waves. Using this method, we find a conserving pseudoscalar, which is equivalent to Lipkin's zilch \cite{Lipkin1964} in antiferromagnetic materials, and which we propose as a measure of chirality for spin-wave excitations.

In order to observe spin-wave chirality related effects, the chiral symmetry inside the material itself must be broken. One possibility for such symmetry breaking is to consider antiferromagnets  with nonzero Lifshitz invariants \cite{Benfatto2006,Udvardi2009,Odashima2013}.  Another way, proposed in this Letter, is to inject a pure spin current, which lifts the $\mathcal{P}$-symmetry, at the same time, keeping the $\mathcal{T}$-symmetry unbroken. As we discuss below, the effect of spin current in the linear regime can be effectively expressed in terms of induced Lifshitz invariants in the spin-wave energy. We demonstrate that in such antiferromagnets with spin current driven chirality, spin wave chirality plays a role similar to electromagnetic chirality  \cite{Tang2010} determining the asymmetry in the spin-wave energy absorption rate with respect to spin current direction. We also show that on a quantum level our spin-wave chiralty is proportional to the difference between left and right polarized magnon numbers, and propose a linear response theory for the spin-current induced non-equilibrium magnon chirality density.

	
\textit{Nongeometric symmetries.} ---  We consider dissipative magnetization dynamics in a uniaxial antiferromagnet described by the semiclassical Landau-Lifshitz-Gilbert equation
\begin{equation} \label{LLG1}
\dot{\bm M}_{i}  = \gamma \bm M_{i} \times \bm H_{i}^{\mathrm{eff}} - \eta \bm M_{i} \times \dot{\bm M}_{i},
\end{equation}
where $ \gamma $ is a gyromagnetic ratio, $\bm{M}_{i}$ denotes the magnetization for the $i$th sublattice ($i = 1,2$), the effective fields $\bm{H}_{i}^{\mathrm{eff}} = -\delta W/\delta \bm{M}_{i}$ are determined by the magnetic energy functional $W$, and $\eta$ is the Gilbert damping coefficient. The energy dissipation is described by the Rayleigh dissipation function
\begin{equation} \label{abs}
\dot W =-\frac{\eta}{\gamma}\int d^3 r \left( \dot{\bm M}_1^2  + \dot{\bm M}_2^2 \right),
\end{equation}
where $ \eta/\gamma > 0 $ \cite{Akhiezer1968}.

In what follows, we consider a general form of the magnetic energy
\begin{equation} \label{energ}
W = \int d^3 r \left[ w_a + \frac{\delta}{2} \bm M_1 \cdot \bm M_2 
+ \frac{\alpha_{ij}}{2} \bm{\nabla} \bm M_{i} \cdot \bm{\nabla} \bm M_{j}
\right],
\end{equation}
where $\delta$ and $\alpha_{ij}$ are the exchange parameters, and $ w_{a} = -(\beta/2)\left[(\bm M_1\cdot \bm n)^2 + (\bm M_2\cdot \bm n)^2 \right] $ corresponds to the uniaxial anisotropy energy density, where $ \bm{n} $ is the unit vector along the anisotropy axis. In what follows, we take $\alpha_{ij} = \alpha$ for $i =j$, and $\alpha_{ij} = \alpha'$ otherwise. For $ \beta > 0 $, $w_{a}$ stabilizes uniform antiferromagnetic ordering with $\bm{M}_1 = -\bm{M}_2$ parallel to $\bm{n}$ \cite{Akhiezer1968}.

In the spin-wave approximation, the equations of motion are linearized by taking $ \bm M_i(t,\bm r) = (-1)^{i+1} M_s \bm n + \bm m_i(t, \bm r) $, where $ M_s $ is the saturation magnetization. Transforming to the momentum space $\bm{m}_{i}(t,\bm{r}) = \int d^{3}p \exp(i\bm p\bm r) \tilde{\bm{m}}_{i}(t,\bm{p})$ and keeping only linear terms in the complex $ \tilde{\bm m} = \tilde{\bm m}_1 + \tilde{\bm m}_2$ and  $ \tilde{\bm l} = \tilde{\bm m}_1 - \tilde{\bm m}_2  $,  we express the equations of motions in the following form
\begin{equation} \label{LLG34}
\begin{array}{l}
\dot{\tilde{\bm{m}}} = -\varepsilon_{l}(\bm p) \bm{n} \times \tilde{\bm{l}} + \eta \bm{n} \times \dot{\tilde{\bm{l}}},\\ 
\dot{\tilde{\bm{l}}} = -\varepsilon_{m}(\bm{p}) \bm{n} \times \tilde{\bm{m}} + \eta \bm{n}\times \dot{\tilde{\bm{m}}},
\end{array}
\end{equation}
where $ \varepsilon_{m}(p) =\gamma M_s(\delta + \beta + (\alpha + \alpha')p^2) $, $ \varepsilon_{l}(p) = \gamma M_s(\beta + (\alpha - \alpha')p^2)$, and $\bm{p}$ is the spin-wave wave vector.

For symmetry analysis of  Eqs.~(\ref{LLG34}),  it is convenient to use an analogue of the Silberstein-Bateman representation of the Maxwell's equations \cite{Bialynicki1996}. For this purpose, we combine $ \tilde{\bm{m}} $ and $ \tilde{\bm{l}} $ into the six-component vector $ \phi(t, \bm p) = (\tilde{\bm{m}}(t, \bm p), \tilde{\bm{l}}(t, \bm p))^{T}$.  The equation of motion for $ \phi(t, \bm p) $ can be written in the matrix form $i \partial_t\phi(t, \bm p) = \mathcal{H} \phi(t, \bm p)$
with 
\begin{equation} \label{Ham}
\mathcal{H} = 
\begin{pmatrix}
0 & -\varepsilon_{l}(p) (\hat{\bm S} \cdot \bm n) \\
-\varepsilon_{m}(p) (\hat{\bm S} \cdot \bm n) & 0
\end{pmatrix},
\end{equation}
where for the symmetry analysis we omitted the damping terms.
We introduce the spin-1 matrices $ (\hat{S}_{\alpha})_{\beta \gamma} = -i \epsilon_{\alpha\beta\gamma}$ where  $\epsilon_{\alpha\beta\gamma} $ is the Levi-Civita symbol ($\alpha,\beta,\gamma=x,y,z$).  Although $ \mathcal{H} $ is not Hermitian, it can be easily symmetrized by applying the momentum-dependent variable change $ \phi = \mathcal{N} \bar\phi  $, where $ \mathcal{N} = \diag(\varepsilon_{m}^{-1/2}, \varepsilon_{l}^{-1/2}) $ \cite{Silveirinha2016,Hassani2017}. After this transformation, the equation of motion acquires a Schroedinger-like form 
\begin{equation} \label{Ham0}
i \partial_t \bar{\phi}(t, \bm p) = \mathcal{H}_{0} \bar{\phi}(t, \bm p),
\end{equation}
where the Hermitian matrix $ \mathcal{H}_0 $ is given by the Cartesian product 
$\mathcal{H}_0 = -\sqrt{\varepsilon_{m}\varepsilon_{l}} \sigma_{1} \otimes (\hat{\bm S} \cdot \bm n)$, where $ \sigma_1 $ is the Pauli matrix.

Equation~(\ref{Ham0}) has the form similar to the Silberstein-Bateman representation of the Maxwell's equations in dispersive medium \cite{Bialynicki1996}. Transformation to the electrodynamics is reached by replacing $ \phi $ with $ \phi_{\mathrm{em}} = (\bm{E}, \bm{B})^{T} $, composed from the electric and magnetic field, and $ \mathcal{H}_{0}$ with $\mathcal{H}_{\mathrm{em}} = -(\sqrt{\varepsilon\mu})^{-1} \sigma_{2} \otimes (\hat{\bm S} \cdot \bm p)$, where $ \varepsilon(\bm p) $ and $ \mu(\bm p) $ are the permittivity and permeability of the medium. Notably, $\mathcal{H}_{0}$ and $\mathcal{H}_{\mathrm{em}}$ share similar algebraic structure. The difference between them is related to their transformation properties under $ \mathcal{T}$ and $\mathcal{P}$ symmetries \footnote{For example, under $\mathcal{T}$-symmetry, $ \phi_{\mathrm{em}} $ transforms as  $\mathcal{T}\phi_{\mathrm{em}} \to \sigma_{3} \phi_{\mathrm{em}}$, since the electric (magnetic) field is $\mathcal{T}$-even ($\mathcal{T}$-odd). In contrast, $ \mathcal{T}\phi \to -\phi $.  This means that to transform from spin-wave dynamics to electrodynamics, one has to replace $ \sigma_1 $ in $\mathcal{H}_{0}$  with $ \sigma_{2} = i \sigma_{1} \sigma_{3} $ to comply with the $\mathcal{T}$- and $\mathcal{P}$-invariance.}.

The analogy between spin-wave dynamics and electrodynamics allows us to generalize the symmetry analysis of Maxwell's equations to antiferromagnetic spin waves.  Similar to electrodynamics \cite{Fushchich1987},  the equations of motion (\ref{Ham0}) are invariant under the eight-dimensional algebra of \emph{nongeometric symmetries} \cite{Fushchich1987}. The basis elements of this algebra are given by $ \mathcal{Q}_1  = i\sigma_{2} \otimes (\hat{\bm S} \cdot \bm n) \hat D  $, $ \mathcal{Q}_2  = \sigma_{1} \otimes \hat I  $, $ \mathcal{Q}_3  = \sigma_{3} \otimes (\hat{\bm S} \cdot \bm n) \hat D  $, $ \mathcal{Q}_4  = i\sigma_{2} \otimes \hat D $, $ \mathcal{Q}_5  = \sigma_{0} \otimes (\hat{\bm S} \cdot \bm n) $, $ \mathcal{Q}_6 = \sigma_{3} \otimes \hat D $, $ \mathcal{Q}_7 = \sigma_{0} \otimes \hat I $, and $ \mathcal{Q}_8 = \sigma_{1} \otimes (\hat{\bm S} \cdot \bm n) $, where $ \hat{D} = 2[(\hat{\bm S} \cdot \bm n_{\perp})^2 - \hat I_{3}n^2_{\perp}]/n^2_{\perp} - (\hat{\bm S} \cdot \bm n)^2$, $ \bm{n}_{\perp} = (n_{1},n_{2}, 0)$, $ \hat I_{3} = \mathrm{diag}(0,0,1) $, $ \sigma_{0} $ and $ \hat{I} $ denote two- and three-dimensional unit matrices respectively \cite{Suppl}. 

Some basis elements have clear interpretation. For example, $\mathcal{Q}_8$, which is proportional to $\mathcal{H} \equiv i\partial_t$, represents the symmetry with respect to taking the time derivative. $ \mathcal{Q}_2 $ plays a role similar to the duality transformations of the electromagnetic field \cite{Calkin1965, Zwanziger1968}. It generates a continuous symmetry transformation $\tilde{\bm m} \to \tilde{\bm m} \cosh \theta + \sqrt{\varepsilon_{l}/\varepsilon_{m}} \tilde{\bm l} \sinh \theta$ and $\tilde{\bm l} \to \tilde{\bm l} \cosh \theta + \sqrt{\varepsilon_{m}/\varepsilon_{l}} \tilde{\bm m} \sinh \theta$ for any real parameter $ \theta $.


\textit{Spin-wave chirality conservation law.} --- From the existence of symmetry transformations, we can establish various conservation laws, which can be conveniently written in terms of bilinear forms 
\begin{equation} \label{EOM}
C_{A} = \frac{1}{2} \int d^3 p \phi^{\dag}(t, \bm p) \rho \mathcal{Q}_A \phi(t, \bm p),
\end{equation}
where $ \rho = (\mathcal{N}^{-1})^{\dag}\mathcal{N}^{-1}$ is the measure that takes into account non-Hermitian character of $\mathcal{H}$ \cite{Proskurin2016}.

Since the rotation symmetry with respect to $ \bm{n} $-direction is unbroken, we can introduce spin-wave chirality conservation, associated with conservation of the operator $p_n \mathcal{Q}_{5}$ in Eq.~(\ref{EOM}), where $p_n= \bm{p} \cdot \bm{n}$ is the spin-wave momentum component  along $ \bm n $.  The explicit form of this conservation law in the momentum space is given by
\begin{equation} \label{Cchi}
C_{\chi} = \frac{i}{2}\int\! d^3 p \left[ \varepsilon_{m}(\bm p) \tilde{\bm m}^* \cdot (\bm p_n \times \tilde{\bm m}) + \varepsilon_{l}(\bm p) \tilde{\bm l}^* \cdot (\bm p_n \times \tilde{\bm l}) 
\right].
\end{equation}
This relation is an analog of the Lipkin's zilch \cite{Lipkin1964}. The corresponding real space spin-wave chirality density can be written as 
\begin{equation}\label{Jchi}
 \rho_{\chi}(t, \bm{r}) = \frac{1}{2} \left ( \dot{\bm l} \cdot \nabla_{n} \bm m + \dot{\bm m} \cdot \nabla_{n} \bm l\right ),
\end{equation} 
where $ \nabla_n = \bm n\cdot \bm \nabla $.
  In this case, the total chirality is obtained by taking the volume integral $C_{\chi}=\int d^3 r \rho_{\chi}(t, \bm{r})$.
  
  In order to clarify the physical meaning of the  spin-wave chirality in Eqs.~(\ref{Cchi}, \ref{Jchi}), we rewrite these equations in terms of magnon operators. By applying the Holstein-Primakoff transformation \cite{Rezende1976} for sublattice magnetizations $ M_{1}^{(+)} = \sqrt{2M_s} a $, $ M_{1}^{(-)} = \sqrt{2M_s} a^{\dag} $, $ M_{1}^{z} =  M_s - a^{\dag}a $ and $  M_{2}^{(+)} = \sqrt{2M_s} b^{\dag} $, $ M_{2}^{(-)} = \sqrt{2M_s} b $, $ M_{2}^{z} = -M_s + b^{\dag}b $, where $ a $ and $ b $ are bosonic operators, combined with the Bogolyubov's rotation $ a_{\bm{p}} =  a_{\mathrm{L}\bm{p}}^{\phantom{\dag}}\cosh\theta -  a_{\mathrm{R}-\bm{p}}^{\dag}\sinh\theta $, and $ b_{-\bm{p}}^{\dag} = a_{\mathrm{R}-\bm{p}}^{\dag}\cosh\theta-a_{\mathrm{L}\bm{p}}^{\phantom{\dag}}\sinh\theta $ with $ \tanh\theta = (\varepsilon_{m} - \varepsilon_{l})/(\varepsilon_{m} + \varepsilon_{l} + 2\sqrt{\varepsilon_{m} \varepsilon_{l}}) $,  the total magnon Hamiltonian can be written in terms of left (L) and right (R) polarized magnon number operators
  \begin{equation}\label{HH0}
  \hat{H} =  \frac{1}{2} \sum_{\bm{p}} \omega_{\bm p} \left( a_{\mathrm{L}\bm{p}}^{\dag}a_{\mathrm{L}\bm{p}}^{\phantom{\dag}} + a_{\mathrm{R}\bm{p}}^{\dag}a_{\mathrm{R}\bm{p}}^{\phantom{\dag}} \right),
  \end{equation}
  where the energy dispersion $\omega_{\bm p} = \sqrt{\varepsilon_{l}(\bm p)\varepsilon_{m}(\bm p)}$ is doubly degenerated with respect to polarization directions \cite{Akhiezer1968}.  
  In terms of $a_{\mathrm{L}\bm{p}}$ and $a_{\mathrm{R}\bm{p}}$, $ C_{\chi} $ is determined by the difference in numbers of L- and R-polarized magnons
  \begin{equation}
  \hat{C}_{\chi} = 2 \sum_{\bm{p}} p_n \omega_{\bm p}  \left(a_{\mathrm{L}\bm{p}}^{\dag}a_{\mathrm{L}\bm{p}}^{\phantom{\dag}} - a_{\mathrm{R}\bm{p}}^{\dag}a_{\mathrm{R}\bm{p}}^{\phantom{\dag}} \right). 
  \end{equation} 
 Similar expression for optical helicity and Lipkin's zilch in terms of the  photon numbers is known for a long time \cite{Calkin1965,Connell1965,Zwanziger1968,Afanasiev1996,Trueba1996,Coles2012}.

\textit{Chiral symmetry breaking.} --- At this point, we have established a chirality conservation law for spin waves in antiferromagnets. We now discuss consequences and potential for observation and application.

\begin{figure}
	\centerline{\includegraphics[width=0.39\textwidth]{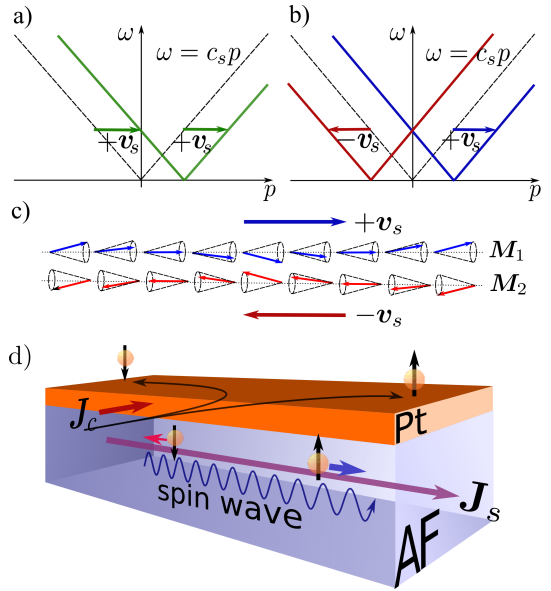}}
	\caption{(color online) a) Schematic picture of the Doppler shift for spin waves with the energy dispersion $\omega = c_{s}p$. Both L- and R-polarized magnon modes shift in the same direction. b) Antiparallel  Doppler shifts for L- and R-modes lift the degeneracy with respect to polarization. c) Magnetizations $ \bm M_{1} $ and $ \bm M_{2} $ boosted in the opposite directions by spin current injection along $ \bm n $; d) Possible experimental realization: charge current $J_{c}$ is converted into pure spin current $ J_{s} $ in the material with large $\theta_{\mathrm{SH}}$ (Pt) and injected into the antiferromagnet (AF) across the interface.}
	\label{fig1}
\end{figure}

We note that $C_{\chi}$ is odd under both transformations, $\mathcal{P}$ and exchange of sublattices $ \bm{m}_{1} \leftrightarrow \bm{m}_{2} $ ($\mathcal{M}$). Therefore, to observe spin-wave chirality related effects, these symmetries should be broken inside the material.
To break the inversion symmetry, we may try to exploit the Doppler shift of spin waves, which is formally reached by replacement $ \partial_t \to \partial_t - \bm{v}_{s} \cdot \bm \nabla $ in the equations of motion, where $ \bm{v}_s $ is the velocity of the moving frame \cite{Swaving2011}.
This effect was observed in ferromagnetic metals under applied spin-polarized current \cite{Vlaminck2008}, and was proposed for antiferromagnets \cite{Swaving2011}. However, pure Doppler shift does not lift the degeneracy between L- and R-polarized modes \cite{Yamane2016}, and, therefore, cannot induce chirality (see Fig.~\ref{fig1}a).

To create chirality, we propose to realize two different Doppler shifts for L- and R-polarized magnons in the \emph{opposite directions}, as schematically shown in  Fig~\ref{fig1}b, which also breaks $\mathcal{M}$-symmetry. Below, we consider how this situation can be experimentally realized. Here, we only note that, formally, this can be achieved by two antiparallel Galilean boosts for $\bm{M}_{1}$ and $\bm{M}_{2}$ sublattice magnetizations (see Fig~\ref{fig1}c), which correspond to the transformation $ \partial_t \to \partial_t \mp v_s \nabla_{n}$ in Eq.~(\ref{LLG1}), where upper (lower) sign is for $ \bm{M}_1 $ ($ \bm{M}_2 $), and we take $ v_{s} $ parallel to $ \bm{n} $.

Applying the transformation $ \partial_t \to \partial_t \mp v_s \nabla_{n}$ to the energy absorption rate in Eq.~(\ref{abs}), we find that for spin-waves traveling in such medium, $ \dot{W} $ splits into symmetric and asymmetric parts under $\mathcal{P}$ and $\mathcal{M}$. The latter part is proportional to the spin-wave chirality
\begin{equation} \label{Wa}
\dot{W}_{\chi} = \frac{2\eta v_s}{\gamma}\!\int\!\! d^3 r \left( \dot{\bm{m}}_1 \cdot \nabla_n \bm{m}_1 - \dot{\bm{m}}_2 \cdot \nabla_n \bm{m}_2 \right) = \frac{2\eta v_s}{\gamma} C_{\chi}.
\end{equation}
This result is the magnetic counterpart of the effect first demonstrated  in optics by Tang and Cohen \cite{Tang2010}.

\textit{Experimental realization.} --- How experimentally realize the antiparallel Galilean boosts for $\bm{M}_{1}$ and  $\bm{M}_{2}$? For this purpose, we invoke the spin-transfer torque (STT) mechanism \cite{Daniels2015}.

We consider pure spin current injected into the antiferromagnet along the $\bm{n}$-direction (see Fig.~\ref{fig1}d). The existence of spin current means that a portion of $\bm{s}_{\uparrow}$-electrons is flowing into the positive direction with the velocity $v_s$, while an equal amount of $\bm{s}_{\downarrow}$-electrons drifts in the opposite directions with $-v_{s}$. If the entire antiferromagnet is in the \emph{exchange dominant} regime \cite{Yamane2016}, intersublattice electron dynamics can be neglected and $\bm{s}_{\uparrow}$ ($\bm{s}_{\downarrow}$) electrons couple only to the $ \bm M_1 $ ($ \bm M_2 $) sublattice. In this case, these two sorts of electrons are able to produce an adiabatic STT onto $\bm{M}_{1}$ and  $\bm{M}_{2}$ pointing in the opposite directions via the Zhang-Li mechanism \cite{Zhang2004}. In particular, $ \bm{s}_{\uparrow} $-electron flow produces the torque $ \bm{\mathfrak{T}}_{1} = M_{s}^{-2}\bm{M}_1 \times \bm{M}_{1} \times (\bm{v}_s\cdot \bm{\nabla}) \bm{M}_1$ acting on $ \bm{M}_{1} $, while at the same time  $ \bm{s}_{\downarrow} $-electrons produce $ \bm{\mathfrak{T}}_{2} = - M_{s}^{-2}\bm{M}_2 \times \bm{M}_{2} \times (\bm{v}_s\cdot \bm{\nabla}) \bm{M}_2$ acting on $ \bm{M}_{2} $, where $\bm v_{s} = \mu_{B} \bm j_{s}/(eM_{s})$ is proportional to the spin current density $\bm{j}_{s}$ (in electric units).

 The spin current can be either injected from the metal with large spin-Hall angle $\theta_{\mathrm{SH}}$ ($\approx 0.1$ in Pt), or created inside metallic antiferromagnet with intrinsic spin-Hall effect  (e.g. $\theta_{\mathrm{SH}} \approx 0.06$ in PtMn \cite{Zhang2014}). To estimate $v_{s}$, we take $j_{s} = \theta_{\mathrm{SH}}j_{c}$ with $j_{c} = 10^{12}$~A/m$^{2}$ being the charge current density previously used to observe STT effects in ferromagnets \cite{Vlaminck2008,Yamaguchi2004}. For $\theta_{\mathrm{SH}} = 0.1$, and $M_{s} = 3.5 \times 10^{5}$ A/m, we obtain $v_{s} = 33$~m/s. We use this value below.

The effect of $\bm{\mathfrak{T}}_{1}$ and $\bm{\mathfrak{T}}_{2}$ on the spin-wave spectrum is equivalent the Doppler shifts of L- and R-polarized modes in the opposite directions, as schematically shown in Fig.\ref{fig1}b \cite{Suppl}. Spin current injection lifts the degeneracy with respect to helicity and turns the antiferromagnet into chiral material with magnonic optical activity and circular dichroism. The characteristic length scale of the dichroism in a typical antiferromagnetic insulator with linearly dispersing spin waves $\omega_{\bm{p}}=c_{s}p$ can be estimated as $\ell_{\mathrm{CD}} = c_{s}^2/(\eta v_{s} \omega) \approx 5$~mm, where we take the spin-wave velocity $c_{s} =10$~km/s, $\eta = 10^{-4}$, and frequency $\omega/2\pi = 1$~THz.

In the spin-wave approximation, the effect of adiabatic STT can be effectively described by  the following Lifshitz invariants in the spin-wave energy
\begin{equation}\label{Hint}
H_{\mathrm{DM}} = \frac{v_{s}}{2}\int d^{3}r \left[ \bm{m}_{1} \cdot (\bm{\nabla}_{n} \times \bm {m}_{1}) + \bm{m}_{2} \cdot (\bm{\nabla}_{n} \times \bm {m}_{2}) \right].
\end{equation}
In the lattice-model language, this expression corresponds to the monoaxial Dzyaloshinskii-Moriya (DM) interaction between the next nearest neighboring sites with effective strength $D_{\mathrm{eff}} = \hbar^{2} \gamma v_{s}/(M_{s}a^{4})$, which can be estimated as $\hbar v_{s}/a \approx 0.5$~K for $j_{c} = 10^{12}$~A/m$^{2}$, where $a$ is the lattice constant. This situation partly resembles spin-current-induced DM in ferromagnets with Rashba coupling \cite{Kikuchi2016}.  Recently, next-nearest-neighboring DM interactions attracted attention in view of the magnonic Nernst effect in antiferromagnets \cite{Cheng2016a,Zyuzin2016}.

\textit{Linear response.} --- The amount of spin-wave chirality induced by the spin-current, can be estimated using the linear response theory. For this purpose, we treat $H_{\mathrm{DM}}$ as a perturbation to the magnon Hamiltonian in Eq.~(\ref{HH0}). The spin-wave chirality density can be calculated using the Kubo formula \cite{Mahan2013}
\begin{equation}\label{Kubo}
\left\langle \rho_{\chi}\right\rangle = -i \int_{-\infty}^{t} d t' e^{-s(t-t')}
\left \langle \left[\hat{\rho}_{\chi}(t), \hat{H}_{\mathrm{\mathrm{DM}}}(t') \right]\right \rangle,
\end{equation}
where the average is taken with the equilibrium density matrix  $\hat{\rho}_{0} =\exp(-\hat{H}/k_B T)$, $s \to 0^{+}$, and the operators $\hat{\rho}_{\chi}$ and $\hat{H}_{\mathrm{\mathrm{DM}}}$ are obtained from Eqs.~(\ref{Jchi}) and (\ref{Hint}) by the Holstein-Primakoff transformation \cite{Suppl}.

Straightforward calculations show that the total spin current induced chirality at the temperature $T$ is obtained as follows  \cite{Suppl}
\begin{equation}\label{Cind}
C_{\chi} = -2  v_{s} \sum_{\bm p} \frac{\partial n_{\bm p}}{\partial \omega_{\bm p}} p_{n}^{2} [\varepsilon_{m}(\bm p) + \varepsilon_{l}(\bm p)],
\end{equation}
where $ n_{\bm p} = (\exp(\omega_{\bm p}/k_{B}T) - 1)^{-1} $ is the equilibrium magnon distribution. For linearly dispersing  magnons, this expression can be integrated explicitly, providing $C_{\chi} = 4\pi^{2}v_{s}\hbar\Omega_{\mathrm{ex}} (k_{B}T)^{4}/(45c_{s}^{5})$, where $\Omega_{\mathrm{ex}} = \gamma M_{s}\delta$ is the exchange frequency. 

To characterize the asymmetry created by the spin current, we propose to normalize the induced chirality in Eq.~(\ref{Cind}) on the total contribution to $C_{\chi}$ from the magnons with positive chirality at the thermal equilibrium $C_{+} = \sum_{\bm p_n>0} \omega_{\bm p}p_{n}\langle a_{\mathrm{L}\bm{p}}^{\dag}a_{\mathrm{L}\bm{p}}^{\phantom{\dag}} + a_{\mathrm{R}-\bm{p}}^{\dag}a_{\mathrm{R}-\bm{p}}^{\phantom{\dag}} \rangle$. Without spin current, this amount of chirality is compensated by exactly the same number of magnons with negative chirality providing $C_{\chi} = 0$. Therefore, we can introduce a dimensionless parameter $g = C_{\chi}/C_{+}$ that can be interpreted as an amount of degeneracy lifted by the spin current. For magnons with linear dispersion, we can estimate this quantity as \cite{Suppl}
\begin{equation}\label{asym}
g = k \frac{v_{s}}{c_{s}} \frac{\hbar\Omega_{\mathrm{ex}}}{k_{B}T},
\end{equation}
where $k \approx 0.69$. In a typical antiferromagnetic insulator with $c_{s} = 10$~km/s, and $\Omega_{\mathrm{ex}}/2\pi = 10$~THz, we estimate $g \approx 0.3$\% at room temperatures for $j_{c} = 10^{12}$~A/m$^{2}$.

\textit{Summary.} --- The symmetry analysis for spin-wave dynamics in antiferromagnets  has been developed by drawing an analogy with the Maxwell's equations. The conservation law for the spin-wave chirality has been established. This quantity, which is determined by the difference in numbers of left- and right-polarized magnons, is directly related to the Lipkin's zilch in electrodynamics \cite{Lipkin1964}. In this respect, we would like to mention Refs.~\cite{Coles2012,Bliokh2011} relevant to recent discussions of magnon spin current \cite{Rezende2016,Rezende2016a,Cheng2016,Zyuzin2016}. We also note that our symmetry approach has a potential extension to coupled  magneto-optical excitations in antiferromagnets \cite{Satoh2010}. 

Pure spin current in antiferromagnets can lift the degeneracy with respect to polarization. In this situation, spin-wave chirality determines the asymmetry in the spin-wave energy absorption rate, similar to its twin optical effect \cite{Tang2010}. The efficiency of the spin current is determined by the ratio $v_{s}/c_{s}$, which can reach $10^{-3}$ for current densities $j_{c} = 10^{11}\mbox{--}10^{12}$~A/m$^{2}$. Such current densities were previously used to observe the Doppler shift of spin waves in ferromagnets \cite{Vlaminck2008}. The experiments can be probed in thin-film interfaces of antiferromagnetic insulator/non-magnetic metal (for example, spin current generated by the spin-Hall effect across Pt/NiO interface was used in Ref.~\cite{Lin2017}), or in metallic antiferromagnets with intrinsic spin-Hall effect \cite{Zhang2014,Sklenar2016}.

\begin{acknowledgments}
	This work was supported by the Government of the Russian Federation Program 02.A03.21.0006, by the Ministry of Education and Science of the Russian Federation, Grant MK-6230.2016.2, by JSPS KAKENHI Grants No. 25287087 and No. 25220803, by EPSRC Grant No. EP/M024423, by the Russian Foundation for Basic Research (Grant No. 17-52-500131), and by the grant from the Foundation for the Development of Theoretical Physics "BASIS".
\end{acknowledgments}

\bibliography{antiferromagnet}


\appendix
\setcounter{equation}{0}
\newpage
\onecolumngrid

\begin{center}
	\large \textbf{Supplemental material: ``Spin-wave chirality and its manifestations in antiferromagnets''}
\end{center}

\subsection{Invariance algebra}
In order to construct the invariance algebra of nongeometric symmetries for spin-dynamics, we diagonalize $\mathcal{H}_{0}$ in the equation of motion
\begin{equation} \label{Ham01}
i \partial_t \tilde{\phi}(t, \bm p) = \mathcal{H}_{0} \tilde{\phi}(t, \bm p),
\qquad \mathcal{H}_0 = -\sqrt{\varepsilon_{m}\varepsilon_{l}}  \sigma_{1} \otimes (\hat{\bm S} \cdot \bm n),
\end{equation}
by a unitary transformation $ \mathcal{U} = U_{2} \otimes \hat{U}_{\Lambda} $ that combines rotation in the tree-dimensional space 
\begin{equation}
\label{eq:UL}
\hat{U}_{\Lambda} =
\left(
\begin{array}{ccc}
-\dfrac{n_1n_3 + i n_2}{\sqrt{2}n_\perp} & \dfrac{n_1n_3 - i n_2}{\sqrt{2}n_\perp} & n_1 \\
-\dfrac{n_2n_3 - i n_1}{\sqrt{2}n_\perp} & \dfrac{n_2n_3 + i n_1}{\sqrt{2}n_\perp} & n_2 \\
\dfrac{n_\perp}{\sqrt{2}}                 & -\dfrac{n_\perp}{\sqrt{2}}             & n_3
\end{array}
\right),
\end{equation}
where $ n_{\perp} = (n_1^2 + n_2^2)^{1/2} $, which diagonalizes $ (\hat{\bm S} \cdot \bm n)$, with the  $SU(2)$ rotation in the sublattice pseudo-space $U_{2} = (\sigma_{0} + i\sigma_{2})/\sqrt{2}$, leading to the diagonal form  
\begin{equation} \label{Hbar}
\bar{\mathcal{H}} = \mathcal{U}^{\dag} \mathcal{H}_{0} \mathcal{U}=
\sqrt{\varepsilon_{m}(\bm p) \varepsilon_{l}(\bm p)} \sigma_3 \otimes \hat{\Lambda}
=\mathrm{diag}(-\omega_{\bm{p}}, \omega_{\bm{p}}, 0, \omega_{\bm{p}},-\omega_{\bm{p}},0),
\end{equation}
where $ \hat{\Lambda} = \mathrm{diag}(-1,1,0) $. 
Similar to the free electromagnetic field \cite{Fushchich1987}, this form suggests the existence of the eight-dimensional Lie algebra of local symmetry transformations  in the momentums space, $\mathcal{Q}_{A}(\bm{p})$ ($ A = 1,\ldots8 $), which map a solution $\bar \phi(t, \bm{p})$ of the equations of motion (\ref{Ham01}) into another solution $\bar \phi'(t, \bm{p}) = \mathcal{Q}_{A} \bar \phi(t, \bm{p})$. These transformations  can be derived using a method developed in Refs.~\cite{Fushchich1987,Proskurin2016}. 

We can briefly summarize this method as follows. We have to find all the matrices $\mathcal{Q}_{A}(\bm{p})$ commutative with $\mathcal{H}_{0}$. The problem is alleviated in the local frame $ \tilde\phi = (\mathcal{N} \mathcal{U})^{-1} \phi $.
In the local frame, it is straight forward to verify (see e.~g. Ref.~\cite{Proskurin2016}) that all transformations, which preserve the conditions $\bm{n} \cdot \bm{m} = \bm{n} \cdot \bm{l} = 0$, and commute with $ \tilde{\mathcal{H}}_{0} $, have the following eight-parametric form
\begin{equation}
\label{eq:Qa}
\tilde{\cal Q}_A = \left(
\begin{array}{cccccc}
q_{11} &   0    & 0 &   0    & q_{15} & 0 \\
0    & q_{22} & 0 & q_{24} &   0    & 0 \\
0    &   0    & 0 &   0    &   0    & 0 \\
0    & q_{42} & 0 & q_{44} &   0    & 0 \\
q_{51} &   0    & 0 &   0    & q_{55} & 0 \\
0    &   0    & 0 &   0    &   0    & 0
\end{array}
\right).
\end{equation}

The basic elements in the linear space of $ \tilde{\cal Q}_A $ can be chosen as follows. First, we identify the four-parametric commutative subalgebra of diagonal matrices
\begin{equation}
\mathcal{\tilde Q}_2 = -\sigma_{3} \otimes \hat \Lambda^2, \quad
\mathcal{\tilde Q}_5 = \sigma_0 \otimes \hat \Lambda, \quad
\mathcal{\tilde Q}_7 = \sigma_0 \otimes \hat \Lambda^2, \quad
\mathcal{\tilde Q}_8 = -\sigma_{3} \otimes \hat \Lambda.
\end{equation}
Second, it is easy to see that all the non-diagonal matrices can be parametrized in a similar way
\begin{equation}
\mathcal{\tilde Q}_1 = i\sigma_{2} \otimes \hat\Lambda \tilde{D}, \quad
\mathcal{\tilde Q}_3 = \sigma_{1} \otimes \hat \Lambda \tilde{D}, \quad
\mathcal{\tilde Q}_4 = \sigma_{1} \otimes \hat \Lambda^2\tilde{D}, \quad
\mathcal{\tilde Q}_6 = i\sigma_{2} \otimes \hat \Lambda^2\tilde{D},
\end{equation}
where $ \sigma_{0} $ denotes $ 2\times2 $ unit matrix, $ \sigma_{i} $ ($i = 1,2,3$) are the Pauli matrices, and
\begin{equation}
\tilde{D} = 
\begin{pmatrix}
0 & 1 & 0 \\
1 & 0 & 0 \\
0 & 0 & 0
\end{pmatrix}.
\end{equation}
The algebraic structure of $ \tilde{\cal Q}_A $ ($ A = 1,\ldots,8 $) becomes clear if we notice that the matrices $ \bar{D} $, $ i\hat \Lambda \tilde D $, and $ -\hat{\Lambda} $ form the Clifford algebra isomorphic to that of $\sigma_{1} $, $ \sigma_{2} $, and $ \sigma_{3} $. The resulting eight-dimensional algebra of $ \bar{\cal Q}_A $ is isomorphic to $ U(2) \otimes U(2) $ \cite{Fushchich1987}.

Transforming back to the original basis $ \mathcal{Q}_A = \mathcal{U}  \mathcal{\tilde Q}_A \mathcal{U}^{\dag} $, we obtain the basic set  
\begin{align} \label{Q1}
\mathcal{Q}_1 & = \sigma_2 \otimes (\hat{\bm S} \cdot \bm n) \hat D , & 
\mathcal{Q}_2  = \sigma_1 \otimes \hat I , \\
\mathcal{Q}_3 & = \sigma_3 \otimes (\hat{\bm S} \cdot \bm n) \hat D , &
\mathcal{Q}_4  = \sigma_2 \otimes \hat D ,\\
\mathcal{Q}_5 & = \sigma_0 \otimes (\hat{\bm S} \cdot \bm n)  , & 
\mathcal{Q}_6 = \sigma_3 \otimes \hat D \\
\mathcal{Q}_7 &= \sigma_0 \otimes \hat I, &
\mathcal{Q}_8 = \sigma_1 \otimes (\hat{\bm S} \cdot \bm n),
\label{Q8}
\end{align}
where $ \hat{D} = \hat{U}_{\Lambda} \tilde{D} \hat{U}_{\Lambda}^{\dag}= 2[(\hat{\bm S} \cdot \bm n_{\perp})^2 - \hat I_{3}n^2_{\perp}]/n^2_{\perp} - (\hat{\bm S} \cdot \bm n)^2$, $ \bm{n}_{\perp} = (n_{1},n_{2}, 0)$, and $ \hat I_{3} = \mathrm{diag}(0,0,1)$. 

All the matrices $\mathcal{Q}_{A}$ ($A=1,\ldots8$) commute with $\mathcal{H}_{0} $, and transform a solution of the equations of motion $ \bar \phi(t, \bm p) $, into another solution  $\bar \phi(t,\bm p) \to \bar \phi'(t, \bm p) = \mathcal{Q}_{A} \bar \phi(t, \bm p)$. 

\subsection{Spin-wave chirality quantization}
We now derive a quantum expression for the spin-wave chirality. For this purpose, we apply Holstein-Primakoff transformation to the sublattice magnetizations \cite{Rezende1976,Rezende2016}
\begin{align}
M_{1}^{(+)}(t, \bm{r}) & = \sqrt{2M_s} a       (t, \bm{r}), \quad
M_{1}^{(-)}(t, \bm{r})   = \sqrt{2M_s} a^{\dag}(t, \bm{r}), \quad
M_{1}^{z}  (t, \bm{r})   =  M_s - a^{\dag}(t, \bm{r})a(t, \bm{r}),  \\
M_{2}^{(+)}(t, \bm{r}) & = \sqrt{2M_s} b^{\dag}(t, \bm{r}), \quad
M_{2}^{(-)}(t, \bm{r})   = \sqrt{2M_s} b       (t, \bm{r}), \quad
M_{2}^{z}  (t, \bm{r})   = -M_s + b^{\dag}(t, \bm{r})b(t, \bm{r}),  \\
\end{align}
where $ M_{i}^{(\pm)} = M_{x} \pm iM_{y} $ ($ i = 1, 2 $), and  $ a(t, \bm{r}) $ and 
$ b(t, \bm{r}) $ denote boson operators, which satisfy standard commutation relations
\begin{align}
[a(t, \bm{r}),a^{\dag}(t, \bm{r}')] &= \delta(\bm{r}-\bm{r'}), \quad
[a(t, \bm{r}),a       (t, \bm{r}')] = [a^{\dag}(t, \bm{r}),a^{\dag}(t, \bm{r}')] = 0,  \\
[b(t, \bm{r}),b^{\dag}(t, \bm{r}')] &= \delta(\bm{r}-\bm{r'}), \quad
[b(t, \bm{r}),b       (t, \bm{r}')] = [b^{\dag}(t, \bm{r}),b^{\dag}(t, \bm{r}')] = 0,  \\
[a(t, \bm{r}),b^{\dag}(t, \bm{r}')] &= [a(t, \bm{r}),b(t, \bm{r}')] = 0,
\end{align} 

By applying the Fourier transformation
\begin{equation}
a(t, \bm{r}) = \frac{1}{\sqrt{V}} \sum_{\bm{p}} e^{i \bm{p} \cdot \bm{r}} a_{\bm{p}}, \qquad
b(t, \bm{r}) = \frac{1}{\sqrt{V}} \sum_{\bm{p}} e^{i \bm{p} \cdot \bm{r}} b_{\bm{p}},
\end{equation}
where $ V $ is the total volume of the system, we can express the total energy
\begin{equation}
W = \int d^{3} r \left(w_a + \frac{\delta}{2} \bm M_1 \cdot \bm M_2 
+ \frac{\alpha_{ij}}{2} \bm{\nabla} \bm M_{i} \cdot \bm{\nabla} \bm M_{j} \right),
\end{equation}
in terms the Hamiltonian 
\begin{equation} \label{H0}
\hat{H} = \frac{1}{2} \sum_{\bm{p}} \left\lbrace
(\varepsilon_{m} + \varepsilon_{l}) \left(a_{\bm{p}}^{\dag}a_{\bm{p}} + b_{\bm{p}}^{\dag}b_{\bm{p}}\right)
+ (\varepsilon_{m} - \varepsilon_{l})\left(a_{\bm{p}}b_{-\bm{p}} + a_{\bm{p}}^{\dag}b^{\dag}_{-\bm{p}}\right) \right\rbrace.
\end{equation}
The diagonal form of the Hamiltonian is reached by the Bogolyubov's transformation
\begin{equation} \label{B1}
a_{\bm{p}}        =  u_{\bm{p}} \alpha_{\bm{p}} - v_{\bm{p}} \beta_{-\bm{p}}^{\dag}, \qquad
b_{-\bm{p}}^{\dag} = -v_{\bm{p}} \alpha_{\bm{p}} + u_{\bm{p}} \beta_{-\bm{p}}^{\dag},
\end{equation}
with real parameters 
\begin{equation}
u_{\bm{p}} \equiv \cosh\theta = \frac{\varepsilon_{m}+\varepsilon_{l} + 2\sqrt{\varepsilon_{m}\varepsilon_{l}}}{\left[\left(\varepsilon_{m}+\varepsilon_{l} + 2\sqrt{\varepsilon_{m}\varepsilon_{l}}\right)^2 - \left(\varepsilon_{m}-\varepsilon_{l}\right)^2 \right]^{\frac{1}{2}}}, \quad
v_{\bm{p}} \equiv \sinh\theta = \frac{\varepsilon_{m}-\varepsilon_{l}}{\left[\left(\varepsilon_{m}+\varepsilon_{l} + 2\sqrt{\varepsilon_{m}\varepsilon_{l}}\right)^2 - \left(\varepsilon_{m}-\varepsilon_{l}\right)^2 \right]^{\frac{1}{2}}},
\end{equation}
that satisfy the identity $ u_{\bm{p}}^{2} - v_{\bm{p}}^{2} = 1 $, which warrants standard commutation rules for boson operators $ \alpha_{\bm{p}}(t) $ and $ \beta_{\bm{p}}(t) $. In terms of new operators, the Hamiltonian is given by
\begin{equation} \label{HH}
\hat{H} = \frac{1}{2} \sum_{\bm{p}} \omega_{\bm p} \left( \alpha_{\bm{p}}^{\dag}\alpha_{\bm{p}} + \beta_{\bm{p}}^{\dag}\beta_{\bm{p}} \right),
\end{equation}
where $ \omega_{\bm p} = \sqrt{\varepsilon_{m}(\bm p)\varepsilon_{l}(\bm p)} $.

In order to derive a quantum expression for the spin-wave chirality, we start with the following expression
\begin{equation}
C_{\chi} = M_{s}^{-1}\int d^{3} r \left(\dot{\bm{m}}_{1} \cdot \nabla_{n} \bm{m}_{1} - \dot{\bm{m}}_{2} \cdot \nabla_{n} \bm{m}_{2} \right),
\end{equation} 
and apply the following expansion to the local magnetizations
\begin{align}
\bm{m}_{1}(t, \bm{r}) &= \sqrt{\frac{M_s}{V}} \sum_{\bm{p}} \left(\bm{\mu} a_{\bm{p}}(t) e^{i \bm{p}\cdot\bm{r}} + \bm{\mu}^{*} a_{\bm{p}}^{\dag}(t) e^{-i \bm{p}\cdot\bm{r}} \right), \\
\bm{m}_{2}(t, \bm{r}) &= \sqrt{\frac{M_s}{V}} \sum_{\bm{p}} \left(\bm{\mu}^{*} b_{\bm{p}}(t) e^{i \bm{p}\cdot\bm{r}} + \bm{\mu} b_{\bm{p}}^{\dag}(t) e^{-i \bm{p}\cdot\bm{r}} \right),
\end{align}
where $ \bm{\mu} = (1, -i, 0)/\sqrt{2} $ ($ \bm{\mu}^{*} = (1, i, 0)/\sqrt{2} $) denotes the polarization vector with the helicity $ -1 $ ($ +1 $). Using this expansion, we obtain
\begin{equation}
\hat C_{\chi} = i\sum_{\bm{p}} p_n \left(\dot{a}_{\bm{p}}^{\dag} a_{\bm{p}} -\dot{a}_{\bm{p}} a_{\bm{p}}^{\dag} + \dot{b}_{-\bm{p}}^{\dag} b_{-\bm{p}} -\dot{b}_{-\bm{p}} b_{-\bm{p}}^{\dag} \right).
\end{equation}
By applying the Bogolyubov's transformation in Eq.~(\ref{B1}), the diagonal form is reached
\begin{equation}
\hat C_{\chi} = i\sum_{\bm{p}} p_n \left( \dot{\alpha}_{\bm{p}}^{\dag} \alpha_{\bm{p}} - \alpha_{\bm{p}}^{\dag}  \dot{\alpha}_{\bm{p}}
-\dot{\beta}_{\bm{p}}^{\dag} \beta_{\bm{p}} + \beta_{\bm{p}}^{\dag} \dot{\beta}_{\bm{p}} \right).
\end{equation}
By solving the elementary equations of motions for the boson operators $ \dot{\alpha}(t) = i[\alpha, \hat{H}] $ and $ \dot{\beta}(t) = i[\beta, \hat{H}] $ with $ \hat{H} $ given by Eq.~(\ref{HH}), we obtain
\begin{equation}
\hat C_{\chi} = 2 \sum_{\bm{p}} p_n \omega_{\bm{p}} \left(\hat{N}_{\bm{p}}^{(\mathrm{L})} - \hat{N}_{\bm{p}}^{(\mathrm{R})} \right), 
\end{equation} 
where $ \hat{N}_{\bm{p}}^{(\mathrm{L})} = \alpha_{\bm{p}}^{\dag}\alpha_{\bm{p}} $ and $ \hat{N}_{\bm{p}}^{(\mathrm{R})} = \beta_{\bm{p}}^{\dag}\beta_{\bm{p}} $. Note that the magnon number operator $ \hat{N}_{\bm{p}}^{(\mathrm{L})} $ ($ \hat{N}_{\bm{p}}^{(\mathrm{R})} $) corresponds to the magnon mode with left (right) helicity (see e.g. \cite{Cheng2016a}).


\subsection{Antiferromagnetic spin waves in the presence of a pure spin current}
We consider an antiferromagnet under the \emph{exchange-dominant} approximation \cite{Yamane2016}, where we can neglect intersublattice electron dynamics. In this  ase, spin-majority (spin-minority) electrons couple only to $\bm{M}_{1}$ ($\bm{M}_{2}$) sublattice.  Taking into account that spin-wave dynamics takes place at much longer wave length than that of the electrons, the spin-transfer torque that the electrons exert onto the $i$th sublattice magnetization can be written in the Zhang-Li form \cite{Zhang2004}
\begin{equation}\label{STT}
\mathfrak{T}_{i} = -\frac{1}{M_{s}^{2}} \bm{M}_{i}\times(\bm{M}_{i} \times (\bm{v}_{s} \cdot \bm{\nabla})\bm{M}_{i}) - \frac{\xi}{M_{s}} \bm{M}_{i} \times (\bm{v}_{s} \cdot \bm{\nabla})\bm{M}_{i},
\end{equation}
where the first (second) term is the (non)adiabatic torque, $\bm{v}_{s} = \mu_{B} \bm{j}_{c}^{(i)}/(e M_{s})$ with $\bm{j}_{c}^{(i)}$ being the spin-polarized electron current coupled with the $i$th sublattice, and $ \xi \lesssim 1 $ is a dimensionless parameter \cite{Yamane2016,Zhang2004}.  We imply that a pure electronic  spin current is pumped into the antiferromagnet along the magnetic ordering direction $\bm{n}$ .  This means that the spin-polarized current $\bm{j}_{c}^{(\uparrow)}$ of spin-up electrons is flowing in the positive direction exerting the torque $\bm{\mathfrak{T}_{1}}$, while the current $\bm{j}_{c}^{(\downarrow)} = -\bm{j}_{c}^{(\uparrow)}$ of spin-down electrons is flowing in the negative direction producing $\bm{\mathfrak{T}}_{2} = -\bm{\mathfrak{T}}_{1}$.

The equations of motion for sublattice magnetizations in the presence of the spin torques take the following form 
\begin{align} \label{LLG10}
\dot{\bm M}_1 & = \gamma \bm M_1 \times \bm H_1^{\mathrm{eff}} + \eta \bm M_1 \times \dot{\bm M}_1 -\frac{v_s}{M_{s}^{2}} \bm{M}_{1}\times(\bm{M}_{1} \times \nabla_{n} \bm{M}_{1}) - \frac{\xi v_{s}}{M_{s}} \bm{M}_{1} \times \nabla_{n} \bm{M}_{1},\\ \label{LLG2}
\dot{\bm M}_2 & = \gamma \bm M_2 \times \bm H_2^{\mathrm{eff}} + \eta \bm M_2 \times \dot{\bm M}_2 + \frac{v_{s}}{M_{s}^{2}} \bm{M}_{2}\times (\bm{M}_{2} \times \nabla_{n} \bm{M}_{2}) + \frac{\xi v_{s}}{M_{s}} \bm{M}_{2} \times \nabla_{n} \bm{M}_{2},
\end{align}
where we use $\nabla_{n} = (\bm{n} \cdot \bm{\nabla})$. By linearizing these equations and rewriting them in terms of $\bm{m}$ and $\bm{l}$ in momentum space, we obtain 
\begin{align}
\label{LLG3}
\dot{\bm{m}} - \eta (\bm{n} \times \dot{\bm{l}}) & = -\varepsilon_{l}(p) (\bm{n} \times \bm{l}) - i p_n v_{s} \bm{n}\times(\bm{n} \times \bm{l}) - i\xi p_n v_{s} (\bm{n} \times \bm{m}),    \\
\label{LLG4}
\dot{\bm{l}} - \eta (\bm{n} \times \dot{\bm{m}}) & = -\varepsilon_{m}(p) (\bm{n} \times \bm{m}) - i p_n v_{s} \bm{n}\times(\bm{n} \times \bm{m}) - i\xi p_n v_{s} (\bm{n} \times \bm{l}),
\end{align}
where $ p_n = (\bm n \cdot  \bm p) $.

Note that in the linear approximation, the adiabatic spin transfer torque contribution to the equations of motion is equivalent to the existence of the following Lifshitz invariants in the energy density
\begin{equation} \label{DM}
w_{\mathrm{DM}} = \frac{D}{2} \left( \bm m_1 \cdot (\bm \nabla_n \times \bm m_1) + \bm m_2 \cdot (\bm \nabla_n \times \bm m_2) \right),
\end{equation}
where the parameter $ D $ corresponds to $v_{s}/(\gamma M_s)$.

Rewriting Eqs.~(\ref{LLG3}, \ref{LLG4}) in the matrix form and resolving them with respect to the time derivatives, we obtain
\begin{equation}
i
\begin{pmatrix}
\dot{\bm{m}} \\
\dot{\bm{l}}
\end{pmatrix}
=\frac{1}{1 + \eta^2}
\begin{pmatrix}
i \eta \varepsilon_{m} + i(\eta-\xi)p_n v_{s} (\hat{\bm{S}} \cdot \bm{n}) & 
- p_n v_s (1 + \eta \xi) - \varepsilon_{l} (\hat{\bm{S}} \cdot \bm{n})  \\
- p_n v_s (1 + \eta \xi) - \varepsilon_{m} (\hat{\bm{S}} \cdot \bm{n}) &
i \eta \varepsilon_{l} + i(\eta-\xi)p_n v_{s} (\hat{\bm{S}} \cdot \bm{n})
\end{pmatrix}
\begin{pmatrix}
\bm{m} \\
\bm{l}
\end{pmatrix},
\end{equation}
where we replaced $ (\hat{\bm{S}} \cdot \bm{n})^2 = \hat{I} $, since $\bm{n} \cdot \bm{m} = \bm{n} \cdot \bm{l} = 0$.  The matrix on the right hand side has the following nontrivial eigenvalues
\begin{align}
\lambda_{1} &= -i(\eta - \xi) p_n v_{s} + \frac{i\eta}{2} (\varepsilon_{m} + \varepsilon_{l}) -
\sqrt{\varepsilon_{m} \varepsilon_{l} -\frac{\eta^2}{4} (\varepsilon_{m} - \varepsilon_{l})^2 + (1 + \eta\xi)^2 p_n^2 v_{s}^2 -  (1 + \eta\xi) p_n v_{s}(\varepsilon_{m} + \varepsilon_{l})}, \\
\lambda_{2} &= -i(\eta - \xi) p_n v_{s} + \frac{i\eta}{2} (\varepsilon_{m} + \varepsilon_{l}) +
\sqrt{\varepsilon_{m} \varepsilon_{l} -\frac{\eta^2}{4} (\varepsilon_{m} - \varepsilon_{l})^2 + (1 + \eta\xi)^2 p_n^2 v_{s}^2 -  (1 + \eta\xi) p_n v_{s}(\varepsilon_{m} + \varepsilon_{l})}, \\
\lambda_{3} &= i(\eta - \xi) p_n v_{s} + \frac{i\eta}{2} (\varepsilon_{m} + \varepsilon_{l}) -
\sqrt{\varepsilon_{m} \varepsilon_{l} -\frac{\eta^2}{4} (\varepsilon_{m} - \varepsilon_{l})^2 + (1 + \eta\xi)^2 p_n^2 v_{s}^2 +  (1 + \eta\xi) p_n v_{s}(\varepsilon_{m} + \varepsilon_{l})}, \\
\lambda_{4} &= i(\eta - \xi) p_n v_{s} + \frac{i\eta}{2} (\varepsilon_{m} + \varepsilon_{l}) +
\sqrt{\varepsilon_{m} \varepsilon_{l} -\frac{\eta^2}{4} (\varepsilon_{m} - \varepsilon_{l})^2 + (1 + \eta\xi)^2 p_n^2 v_{s}^2 +  (1 + \eta\xi) p_n v_{s}(\varepsilon_{m} + \varepsilon_{l})}.
\end{align}

In order to clarify the physical meaning of these eigenmodes, let us make the following approximations. (i) We neglect nonadiabatic spin transfer torque ($\xi = 0$) and  keep only leading order terms in $\eta$; (ii) we take the wave vector along $ \bm n $, $ p=p_n $(iii) we approximate $\varepsilon_{m} \approx \gamma M_s \delta$ and $\varepsilon_{l} = \gamma M_s (\alpha-\alpha')p^2$. In this case, for $ p \gg p_s$ the system accommodates two circularly polarized and linearly dispersing eigenmodes with the frequencies
\begin{align} \label{op}
\omega_{\bm{p}}^{(+)} & =  i \eta (\Delta_s - p v_{s}) + c_s |p - p_s|, \\    \label{om}
\omega_{\bm{p}}^{(-)} & =  i \eta (\Delta_s + p v_{s}) + c_s |p + p_s|,
\end{align}
where $ c_{s} = \sqrt{\gamma^2 M_s^2 \delta(\alpha - \alpha')^2 + v_s^2}$ is the spin-wave velocity, $ \Delta_s = \gamma M_s \delta/2 $ is the symmetric part of the spin-wave damping, and $p_s = \gamma M_s v_s \delta/(2 c_s^2)$.

Equations~(\ref{op}, \ref{om}) demonstrate that the effect of a pure spin current on spin-wave propagation is twofold. First, the spin current lifts the degeneracy between of left and right polarized spin waves, which is given by the real parts of $ \omega_s^{(\pm)} $. This effect is analogous to optical activity of electromagnetic waves propagating in gyrotropic medium. Second, it lifts the degeneracy in the damping of these waves that can be found from the imaginary parts of $ \omega_s^{(\pm)} $, which can be considered as a circular dichroism of spin waves. The characteristic length scale of the dichroism can be estimated as $\ell_{\mathrm{CD}} = c_{s}/(\eta v_{s} p) = c_{s}^{2}/(\eta v_{s} \omega)$.


\subsection{Linear response theory}

We consider the emergence of spin-wave chirality density $\rho_{\chi}(\bm{r})$ as a response to spin current flow. The total magnon Hamiltonian in the presence of the spin-current becomes $\hat{H}_{t} = \hat{H} + \hat{H}_{\mathrm{DM}}$, where $ \hat{H} $ is given by Eq.~(\ref{HH}) and  $ \hat{H}_{\mathrm{DM}} $ takes into account the effect of adiabatic spin-transfer torques in Esq.~(\ref{LLG3}, \ref{LLG4})
\begin{equation}\label{Hint0}
\hat{H}_{\mathrm{DM}} = \frac{1}{2} \int d^3 r v_s(\bm{r}) \left[ \bm{m}_1 \cdot (\bm{\nabla}_n \times \bm {m}_1) + \bm{m}_2 \cdot (\bm{\nabla}_n \times \bm {m}_2)\right].
\end{equation}
Note that we have explicitly broken the translational symmetry by assuming $ \bm{r} $-dependence in $ v_{s} $ to carry out all the calculations at finite momentum $ \bm{q} $, taking $ \bm q \to 0 $ limit at the end.

By applying the Holstein-Primakoff transformation, we obtain
\begin{equation}\label{Hint1}
\hat{H}_{\mathrm{DM}} = \sum_{\bm p \bm q} \left(p_n + \frac{q_n}{2}\right) v_{s}(\bm q)
\left[ b^{\dag}_{\bm p + \bm q} b_{\bm p} - a^{\dag}_{\bm p + \bm q} a_{\bm p} \right].
\end{equation}
Under the Bogolyubov's transformations in Eqs.~(\ref{B1}), this expression is transformed into
\begin{multline}\label{Hint2}
\hat{H}_{\mathrm{DM}} = -\sum_{\bm p \bm q} \left(p_n + \frac{q_n}{2}\right) v_{s}(\bm q)
\left[
\left(u_{\bm p + \bm q} u_{\bm p} + v_{\bm p + \bm q} v_{\bm p}\right)
\left ( \alpha_{\bm p + \bm q}^{\dag} \alpha_{\bm p} + \beta_{-\bm p}^{\dag} \beta_{-\bm p -\bm q}\right )
\right.
\\
\left.
-\left(u_{\bm p + \bm q} v_{\bm p} + v_{\bm p + \bm q} u_{\bm p}\right)
\left ( \alpha_{\bm p + \bm q}^{\dag} \beta_{-\bm p}^{\dag} + \alpha_{\bm p} \beta_{-\bm p -\bm q}\right )
\right].
\end{multline}

The linear response spin-wave chirality density created by spin-current is given by the Kubo formula \cite{Mahan2013}
\begin{equation}\label{Kubo1}
\left\langle \rho_{\chi}(\bm q) \right\rangle = -i \int_{-\infty}^{t} d t' e^{-s(t-t')}
\left \langle \left[\hat{\rho}_{\chi}(t,\bm q), \hat{H}_{\mathrm{int}}(t') \right]\right \rangle,
\qquad s \to 0^{+},
\end{equation}
where $ \langle \dots \rangle$ means the average with equilibrium density matrix $ \hat{\rho}_{0} = \exp(-\hat{H}/k_B T) $, and the chirality density operator 
$ \hat{\rho}_{\chi}(t,\bm q) = \exp(\hat{H}t) \hat{\rho}_{\chi}(\bm q) \exp(-\hat{H}t) $ is defined as
\begin{multline} \label{rho1}
\hat{\rho}_{\chi}(\bm q) = \frac{1}{2}\int d^3 r e^{-i\bm q \cdot \bm r} \left( \dot{\bm l} \cdot \nabla_{n} \bm m + \dot{\bm m} \cdot \nabla_{n} \bm l \right)
=\int d^3 r e^{-i\bm q \cdot \bm r} \left( \dot{\bm m_{1}} \cdot \nabla_{n} \bm m_{1} - \dot{\bm m_{2}} \cdot \nabla_{n} \bm m_{2} \right)\\
=\sum_{\bm p} i p_n \left( \dot{a}^{\dag}_{\bm p - \bm q} a_{\bm p} - \dot{a}_{\bm p + \bm q} a^{\dag}_{\bm p} + \dot{b}^{\dag}_{-\bm p - \bm q} b_{-\bm p} - \dot{b}_{-\bm p + \bm q} b^{\dag}_{\bm p}  \right) \\
= -  \sum_{\bm p} \left[ (\omega_{\bm p + \bm q} + \omega_{\bm p})p_n + q_n \omega_{\bm p} \right] \left(u_{\bm p + \bm q} u_{\bm p} - v_{\bm p + \bm q} v_{\bm p}\right) \left(\alpha_{\bm p}^{\dag}\alpha_{\bm p + \bm q} + \beta^{\dag}_{-\bm p - \bm q}\beta_{-\bm p} \right)\\
-   \sum_{\bm p} \left[ (\omega_{\bm p + \bm q} - \omega_{\bm p})p_n - q_n \omega_{\bm p} \right] \left(u_{\bm p + \bm q} v_{\bm p} - v_{\bm p + \bm q} u_{\bm p}\right) \left ( \alpha^{\dag}_{\bm p} \beta^{\dag}_{-\bm p - \bm q}
+ \alpha_{\bm p + \bm q} \beta_{-\bm p} \right ).
\end{multline}

The straightforward calculation in Eq.~(\ref{Kubo1}) with Eqs.~(\ref{Hint2}, \ref{rho1}) gives the following answer
\begin{multline} \label{rho2}
\left\langle \rho_{\chi}(\bm q) \right\rangle = \frac{1}{2}v_{s}(\bm q) \sum_{\bm p}
A_{\bm p, \bm q} \left [
\frac{\langle\hat{N}^{(\mathrm{L})}_{\bm p + \bm q}\rangle - \langle\hat{N}^{(\mathrm{L})}_{\bm p}\rangle}{\omega_{\bm p + \bm q} - \omega_{\bm p} + is} +
\frac{\langle\hat{N}^{(\mathrm{R})}_{-\bm p - \bm q}\rangle - \langle\hat{N}^{(\mathrm{R})}_{-\bm p}\rangle}{\omega_{\bm p + \bm q} - \omega_{\bm p} - is} 
\right ]\\
+\frac{1}{2}v_{s}(\bm q) \sum_{\bm p}
\left [
\frac{B_{\bm p, \bm q}}{\omega_{\bm k + \bm q} + \omega_{\bm p} + is} +
\frac{B_{\bm p, \bm q}}{\omega_{\bm k + \bm q} + \omega_{\bm p} - is} 
\right ],
\end{multline}
where 
\begin{align}
A_{\bm p, \bm q} & = \left(2p_n + q_n\right)\left[ (\omega_{\bm p + \bm q} + \omega_{\bm p})p_n + q_n \omega_{\bm p} \right] \left(v_{\bm p + \bm q}^{2} v_{\bm p}^{2} - u_{\bm p + \bm q}^{2} u_{\bm p}^{2}\right),  \\
B_{\bm p, \bm q} & = \left(2p_n + q_n\right)\left[ (\omega_{\bm p + \bm q} - \omega_{\bm p})p_n - q_n \omega_{\bm p} \right] \left(u_{\bm p + \bm q}^{2} v_{\bm p}^{2} - v_{\bm p + \bm q}^{2} u_{\bm p}^{2}\right).
\end{align}

Taking the real part in Eq.~(\ref{rho2}) with $ s \to 0^{+} $, and noting that in thermodynamic equilibrium 
$ \langle\hat{N}^{(\mathrm{L})}_{\bm p}\rangle = \langle\hat{N}^{(\mathrm{R})}_{\bm p}\rangle = n_{\bm{p}} $, where
$ n_{\bm p} = \left[\exp(\omega_{\bm p}/k_B T) - 1\right]^{-1} $ is the magnon Bose-Einstein distribution function, we obtain 
\begin{equation}
\left\langle \rho_{\chi}(\bm q) \right\rangle =  \lambda_{\chi}(\bm q) v_{s}(\bm q),
\end{equation}
where the susceptibility is defined as
\begin{equation}
\lambda_{\chi}(\bm q) = \sum_{\bm p} \left (  \frac{A_{\bm p,\bm q}(n_{\bm p + \bm q} - n_{\bm p})}{\omega_{\bm p + \bm q} - \omega_{\bm p}} + \frac{B_{\bm p,\bm q}}{\omega_{\bm p + \bm q} + \omega_{\bm p}} \right ).
\end{equation}
In the $ \bm q \to 0 $ limit, $ A_{\bm p, \bm q \to 0} = p_{n}^{2} [\varepsilon_{m}(\bm p) + \varepsilon_{l}(\bm p)] $ and $ B_{\bm p, \bm q \to 0} = 0 $, and we obtain
\begin{equation}
\lambda_{\chi}(\bm q \to 0) = -2 \sum_{\bm p} \frac{\partial n_{\bm p}}{\partial \omega_{\bm p}} p_{n}^{2} [\varepsilon_{m}(\bm p) + \varepsilon_{l}(\bm p)],
\end{equation}
which gives 
\begin{equation}\label{CC}
C_{\chi} = -2  v_{s} \sum_{\bm p} \frac{\partial n_{\bm p}}{\partial \omega_{\bm p}} p_{n}^{2} [\varepsilon_{m}(\bm p) + \varepsilon_{l}(\bm p)],
\end{equation}

\subsection{Asymmetry factor}

To define the asymmetry factor, we normalize current induced $C_{\chi}$ in Eq.~(\ref{CC}) on the total chirality of magnons with negative chirality  at the thermodynamic equillibrium, which is defined as a sum chirality of R-polarized magnons moving in the positive direction with respect to the rotation axis $\bm n$ and chirality of L-polarized magnons moving in the negative direction
\begin{equation}\label{CP}
C_{-} =  \sum_{\bm p_{n}<0} \omega_{\bm p} p_{n} \langle N_{\bm p}^{\mathrm{L}} \rangle - \sum_{\bm p_{n}>0} \omega_{\bm p} p_{n} \langle N_{\bm p}^{\mathrm{R}}\rangle  = -\sum_{\bm p_{n}>0} \omega_{\bm p} p_{n} \left ( \langle N_{\bm p}^{\mathrm{R}} \rangle + \langle N_{-\bm p}^{\mathrm{L}} \rangle \right ) = -2 \sum_{\bm p} \omega_{\bm p} |p_{n}| n_{\bm p},
\end{equation}
where $ \langle \dots \rangle$ means the thermodynamic average. 

For magnons with linear dispersion $ \omega_{\bm p} = c_{s}p $, we can estimate $C_{\chi}$ and $ C_{-} $ by replacing the summation in Eqs.~(\ref{CC}, \ref{CP}) by integration in the infinite limits
\begin{equation}
C_{\chi} = -2v_{s}\sum_{\bm p}\frac{\partial n_{\bm p}}{\partial \omega_{\bm p}} p_{n} \left[\varepsilon_{m}(\bm p) + \varepsilon_{l}\right(\bm p) ] = -\frac{v_{s}\hbar\Omega_{\mathrm{ex}}}{3\pi^{2}c_{s}^{5}} \int_{0}^{\infty} d \omega\, \omega^{4} \frac{\partial n}{\partial \omega}=\frac{4\pi^{2}v_{s}\hbar\Omega_{\mathrm{ex}}}{45 c_{s}^{5}} (k_{B}T)^{4},
\end{equation}
\begin{equation}
C_{-} = -2 \sum_{\bm p} |p_{n}| \omega_{\bm p} n_{\bm p} = -\frac{(k_{B}T)^{5}}{2\pi^{2}c_{s}^{4}}\int_{0}^{\infty}\frac{d x\, x^{4}}{e^{x} - 1} = -\frac{12 \zeta(5)}{\pi^{2}c_{s}^{4}}(k_{B}T)^{5},
\end{equation}
where $ \Omega_{\mathrm{ex}} = \gamma M_{s} \delta $ is the exchange frequency.  Therefore, the asymmetry factor is calculated as 
\begin{equation}
g = -\frac{C_{\chi}}{C_{-}} =\frac{\pi^{4}v_{s}}{135\zeta(5)c_{s}}\frac{\hbar\Omega_{\mathrm{ex}}}{k_{B}T}\approx 0.69 \frac{v_{s}}{c_{s}} \frac{\hbar\Omega_{\mathrm{ex}}}{k_{B}T}.
\end{equation}


\end{document}